\documentclass{article}
\PassOptionsToPackage{numbers,compress}{natbib}
\usepackage[preprint]{neurips_2026}

\usepackage[utf8]{inputenc}
\usepackage[T1]{fontenc}

\usepackage{amsmath}
\usepackage{amssymb}
\usepackage{microtype}
\usepackage{hyperref}
\usepackage{url}
\usepackage{booktabs}
\usepackage{array}
\usepackage{graphicx}
\usepackage{xcolor}
\usepackage{colortbl}
\usepackage{enumitem}
\usepackage{multirow}
\usepackage{makecell}
\usepackage{subcaption}
\usepackage{caption}

\definecolor{darkblue}{rgb}{0, 0, 0.5}
\definecolor{tablegray}{HTML}{F5F6F8}
\hypersetup{colorlinks=true, citecolor=darkblue, linkcolor=darkblue, urlcolor=darkblue}

\newcommand{\bench}{Claw-Eval-Live}
\newcommand{\ntasks}{105}
\newcommand{\ncandidates}{157}

\newcommand{\nmodels}{13}
\newcommand{\nservices}{18}
\newcommand{\ncats}{22}

\newcommand{\passthresh}{0.80}
\newcommand{\nserviceworkflows}{87}
\newcommand{\nworkspace}{18}

\title{\bench: A Live Agent Benchmark for Evolving Real-World Workflows}

\author{%
Chenxin Li$^{1,\dagger}$ \quad Zhengyang Tang$^{2}$ \quad Mingxin Huang$^{3}$ \quad Yunlong Lin$^{4}$ \\
\textbf{Shijue Huang$^{5}$ \quad Shengyuan Liu$^{1}$ \quad Bowen Ye$^{6}$ \quad Rang Li$^{6}$} \\
\textbf{Lei Li$^{7}$ \quad Benyou Wang$^{2}$ \quad Yixuan Yuan$^{1}$} \\
{\normalfont $^{1}$The Chinese University of Hong Kong} \\
{\normalfont $^{2}$The Chinese University of Hong Kong, Shenzhen} \\
{\normalfont $^{3}$South China University of Technology \quad $^{4}$Xiamen University \quad $^{6}$Peking University} \\
{\normalfont $^{5}$The Hong Kong University of Science and Technology \quad $^{7}$The University of Hong Kong} \\
{\scriptsize $\dagger$ Project lead} \\
{\footnotesize Project Page: \href{https://claw-eval-live.github.io}{\texttt{https://claw-eval-live.github.io}}}
}

\begin{document}
\maketitle

\begin{abstract}
LLM agents are expected to complete end-to-end units of work across software tools, business services, and local workspaces.
Yet many agent benchmarks freeze a curated task set at release time and grade mainly the final response, making it difficult to evaluate agents against evolving workflow demand or verify whether a task was executed.
We introduce \bench{}, a live benchmark for workflow agents that separates a refreshable signal layer, updated across releases from public workflow-demand signals, from a reproducible, time-stamped release snapshot.
On the construction side, each release is constructed from public workflow-demand signals, with ClawHub Top-500 skills used in the current release, and materialized as controlled tasks with fixed fixtures, services, workspaces, and graders.
On the grading side, \bench{} records execution traces, audit logs, service state, and post-run workspace artifacts, using deterministic checks when evidence is sufficient and structured LLM judging only for semantic dimensions.
The release contains \ntasks{} tasks spanning controlled business services and local workspace repair, and evaluates \nmodels{} frontier models under a shared public pass rule.
Experiments on these public models reveal that reliable workflow automation remains far from solved, with the leading model passing only 66.7\% of tasks and no model reaching 70\%.
They show that failures are structured by task family and execution surface, with HR, management, and multi-system business workflows as persistent bottlenecks and local workspace repair comparatively easier but unsaturated.
Finally, leaderboard rank alone is insufficient because models with similar pass rates can diverge in overall completion, and task-level discrimination concentrates in a middle band of evaluation tasks.
\bench{} suggests that workflow-agent evaluation should be grounded twice, in fresh external demand and in verifiable agent action.
\end{abstract}

\section{Introduction}
\label{sec:intro}

As Large Language Model (LLM) agents move from single-turn question answering to multi-step execution, the target of evaluation is changing as well.
Prior work has connected LLMs to tool use, planning, and agentic scaffolds~\cite{yao2022react,schick2023toolformer,shen2023hugginggpt,qiao2023taskweaver,wu2024autogen,hong2024metagpt,wang2024openhands}.
Modern agent harnesses such as Codex~\cite{codex}, Claude Code~\cite{claudecode}, OpenClaw~\cite{openclaw}, and Hermes Agent~\cite{hermesagent} equip models to call tools, navigate workspaces, query services, and coordinate actions across applications.
The central question is increasingly whether an agent can complete an end-to-end unit of work rather than merely produce a plausible response.
This shift is reflected in web, desktop, mobile, software, and professional-agent benchmarks~\cite{zhou2024webarena,xie2024osworld,rawles2024androidworld,jimenez2024swebench,xu2025theagentcompany,ye2026claweval}.
Typical deployments now ask agents to reconcile transactions, prepare a meeting brief from multiple systems, diagnose a broken workspace, coordinate an approval chain, or repair a local development stack.
These are workflows rather than isolated prompts.
They require cross-system retrieval, state-changing writes, artifact-level edits, and auditable evidence of what the agent actually did.

\begin{figure*}[t]
\centering
\includegraphics[width=\textwidth]{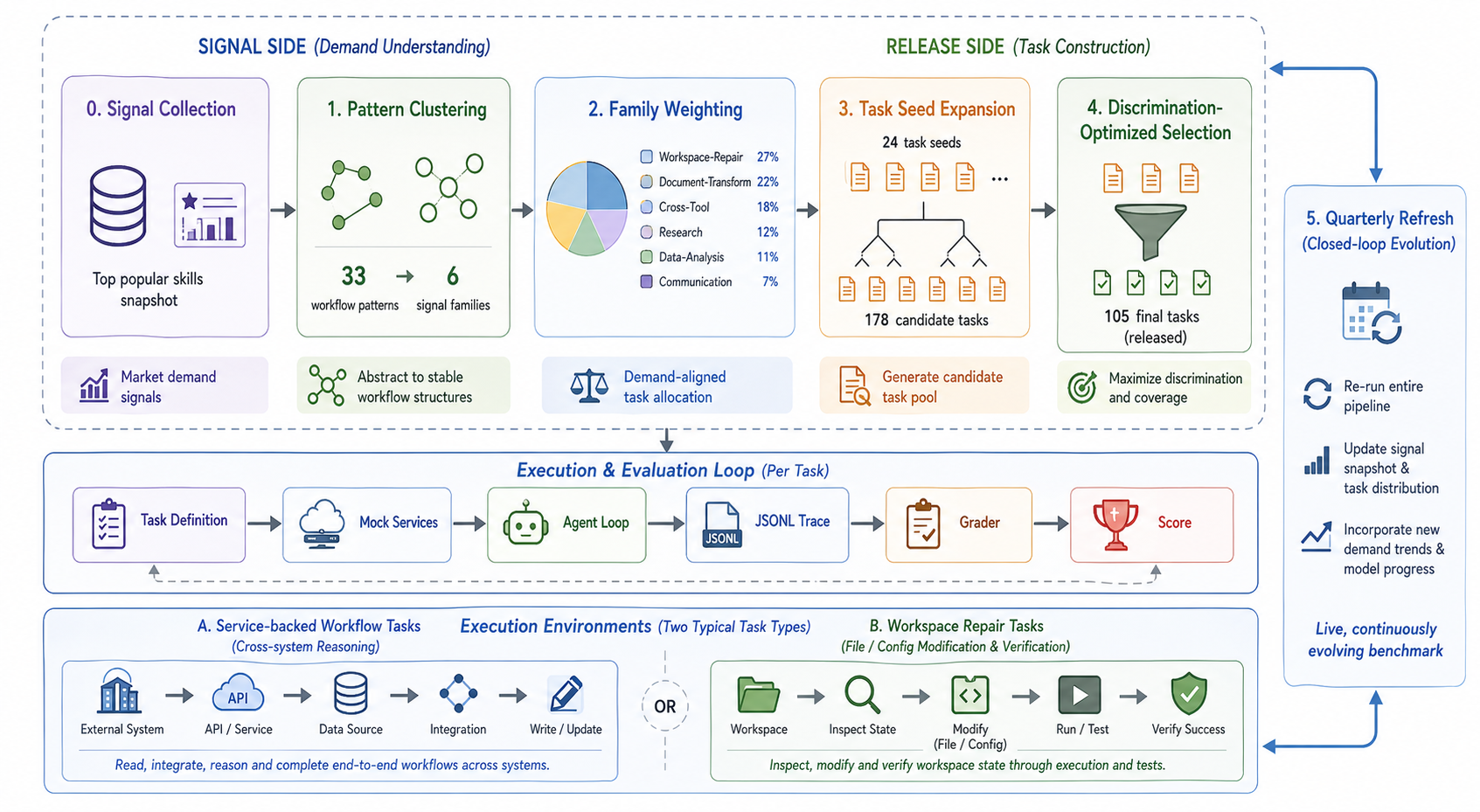}
\caption{Overview of \bench{}.
The benchmark starts from a refreshable snapshot of public workflow signals, clusters and weights demand-side patterns, expands them into candidate tasks, and selects a discrimination-aware public release.
Each released task is executed in a controlled environment, recorded as a trace, and graded from observable evidence.
Quarterly refreshes rerun the pipeline so future releases can track changing demand signals and model progress.}
\label{fig:liveclaw_overview}
\end{figure*}

This creates a problem for benchmark design.
Many public agent benchmarks are still static releases.
Multi-domain suites, web and desktop benchmarks, and software-agent task sets each freeze a particular task mixture at publication time~\cite{agentbench2023,mialon2023gaia,zhou2024webarena,drouin2024workarena,xie2024osworld,jimenez2024swebench,merrill2026terminal}.
Even when these benchmarks are realistic and technically strong, that fixed mixture can become stale.
That matters because the real-world workflow mix keeps changing.
Tool stacks evolve, enterprise bottlenecks shift, and some categories of automation become far more important while others become less central.
A benchmark can therefore remain reproducible while slowly drifting away from the workflows that users currently care about.
For workflow agents, this drift often comes less from any single task than from the task mix itself.
Figure~\ref{fig:liveclaw_overview} summarizes the corresponding design of \bench{}: release construction begins from demand-side public signals, turns them into a time-stamped task snapshot, evaluates each task through observable execution evidence, and reruns the pipeline as signals and models evolve.

Workflow benchmarks face a second problem as well.
A polished memo is not evidence that the agent queried the correct records, compared the right entities, updated the right object, or repaired the broken artifact that the task actually depended on.
This failure mode motivates trajectory-aware and risk-aware evaluation~\cite{ye2026claweval,ruan2023toolemu,yuan2024rjudge,zhang2024agentsafety}.
In workflow settings, the difference between ``sounds right'' and ``did the work'' is not cosmetic.
An agent may produce a fluent report while silently missing a required write, editing the wrong file, or grounding a recommendation in incorrect records.
This matters especially for deployment-facing workflows, which commonly combine service work with local workspace work.
In service-backed tasks, the agent must query records, cross-reference systems, and commit state-changing actions such as drafts, calendar updates, or task creation.
In workspace tasks, the agent must inspect logs, edit files, run commands, and prove that a repair actually worked.
Many high-value workflows cross these boundaries.
Existing benchmarks usually make one side visible: web and device benchmarks emphasize interaction environments, while software and command-line benchmarks emphasize workspace execution~\cite{yoran2024assistantbench,pan2024webcanvas,clawbench2026,xie2024osworld,rawles2024androidworld,jimenez2024swebench,liu2023repobench,merrill2026terminal}.
A benchmark that only scores final responses, or that covers only one of these settings, cannot reliably tell whether the agent actually retrieved the right information, changed the right state, and repaired the right artifact.
For deployment-facing workflows, the missing requirement is their combination.

\bench{} is built around the idea that workflow-agent evaluation should be calibrated twice.
First, the released task distribution should stay close to evolving real-world workflows rather than to a one-time author snapshot.
Second, the score of each task should be anchored in observable execution evidence rather than in final-text plausibility alone.
To that end, \bench{} treats each public release as a time-stamped benchmark snapshot built from a rerunnable signal-to-task pipeline over public workflow signals, then evaluates agents in controlled services and sandboxed workspaces with task-specific, action-grounded graders.

Our contributions are threefold:
\begin{itemize}[nosep,leftmargin=1.2em]
\item We present \bench{}, a live benchmark snapshot for workflow agents that sources tasks from public workflow signals and evaluates both service-backed workflows and workspace repair in a single release.
The current snapshot contains \ntasks{} tasks, \nmodels{} public models, and \nservices{} controlled services plus sandboxed workspaces.
\item We define a signal-to-task release pipeline that maps ClawHub Top-500 signals into released benchmark tasks through clustering, weighting, seed expansion, candidate screening, and optimization-based public-subset selection from a screened pool of \ncandidates{} runnable candidates.
\item We introduce action-grounded hybrid grading and report current results showing that even the best model passes only 66.7\% of tasks, that service-backed workflows remain substantially harder than workspace repair, and that several business-critical families remain far from solved.
\end{itemize}

\section{Related Work}
\label{sec:related}

\noindent\textbf{Agent benchmarks.}
Evaluating LLM-based agents has produced benchmarks spanning tool use, web interaction, desktop environments, and professional tasks.
General suites such as AgentBench~\cite{agentbench2023} and GAIA~\cite{mialon2023gaia} test heterogeneous agent capabilities, while WebArena~\cite{zhou2024webarena}, VisualWebArena~\cite{visualwebarena2024}, Mind2Web~\cite{deng2023mind2web}, BrowserGym~\cite{chezelles2024browsergym}, WebCanvas~\cite{pan2024webcanvas}, AssistantBench~\cite{yoran2024assistantbench}, and OSWorld~\cite{xie2024osworld} make browser or desktop interaction central.
Professional and workplace-oriented benchmarks such as WorkArena~\cite{drouin2024workarena} and TheAgentCompany~\cite{xu2025theagentcompany} move closer to deployed work settings.
\bench{} is complementary to this line: its emphasis is not interface realism alone, but a workflow mixture derived from public demand signals and evaluated inside a reproducible release snapshot.

\noindent\textbf{Code and workspace agent benchmarks.}
Tool and code benchmarks provide the closest precedent for the workspace-repair side of \bench{}.
API-Bank~\cite{li2023api}, ToolBench/ToolLLM~\cite{qin2023toollm}, Gorilla~\cite{patil2023gorilla}, MINT~\cite{wang2023mint}, $\tau$-bench~\cite{yao2025tau}, and MCP-Bench~\cite{wang2025mcp} focus on API or tool manipulation.
HumanEval~\cite{chen2021evaluating}, MBPP~\cite{austin2021program}, DS-1000~\cite{lai2023ds}, CRUXEval~\cite{gu2024cruxeval}, RepoBench~\cite{liu2023repobench}, SWE-bench~\cite{jimenez2024swebench}, OpenHands~\cite{wang2024openhands}, and Terminal-Bench~\cite{merrill2026terminal} move from function-level code generation toward repository or command-line execution.
Within the OpenClaw and Claw-style ecosystem, PinchBench~\cite{pinchbench2026}, WildClawBench~\cite{wildclawbench2026}, ClawBench~\cite{clawbench2026}, Claw-Eval~\cite{ye2026claweval}, and ResearchClawBench~\cite{researchclawbench2026} evaluate coding, sandboxed, web, workflow, or research-agent settings.
\bench{} differs by combining service-backed business workflows and local workspace repair under one signal-calibrated release.

\noindent\textbf{Evaluation methodology.}
Output-only grading can miss cases where an agent produces a plausible artifact without faithfully executing the intended workflow.
Claw-Eval~\cite{ye2026claweval} is the closest methodological relative of \bench{} because it emphasizes trajectory-aware evidence, hybrid grading, and multi-dimensional evaluation of agent behavior rather than final response only.
Other work studies complementary reliability and safety gaps: ToolEmu~\cite{ruan2023toolemu}, R-Judge~\cite{yuan2024rjudge}, Agent-SafetyBench~\cite{zhang2024agentsafety}, ST-WebAgentBench~\cite{levy2024stwebagentbench}, and TrickyArena~\cite{ersoy2025trickyarena} focus on risk or safety, while Watch-Every-Step~\cite{xiong2024watch}, GroundingMe~\cite{li2025groundingme}, and reward-hacking analyses~\cite{recentrewardhacking,macdiarmid2025rewardhacking} highlight process-level or grounding failures.
Related freshness-aware work such as LiveCodeBench~\cite{livecodebench2024}, EvoClaw~\cite{evoclaw2026}, WebArena-Verified~\cite{elhattami2025webarenaverified}, and Online-Mind2Web~\cite{xue2025illusion} further shows that benchmark conclusions can change as tasks, environments, or evaluation criteria age.
\bench{} combines these concerns by grounding \emph{what} is measured in fresh public workflow signals and grounding \emph{how} it is scored in observable execution evidence.

\begin{table*}[t]
\centering
\caption{Positioning of representative agent benchmarks through task sourcing, execution mode, interaction properties, release mode, and scale. Check-style columns indicate whether tasks require multi-step interaction and whether execution is isolated; \emph{Partial} denotes limited or scaffold-dependent support.}
\label{tab:benchmark_attributes}
\scriptsize
\setlength{\tabcolsep}{1.5pt}
\begin{tabular}{l l l c c l l}
\toprule
\textbf{Benchmark} & \textbf{Task sourcing} & \textbf{Exec. mode} & \textbf{Multi-step} & \textbf{Isolated exec.} & \textbf{Release mode} & \textbf{Scale} \\
\midrule
WebArena &
Manual pool &
Web sandbox &
${\checkmark}$ &
${\checkmark}$ &
Static release &
812 / 5 sites \\

OSWorld &
Manual pool &
Desktop sandbox &
${\checkmark}$ &
${\checkmark}$ &
Static release &
369 / 9 apps \\

ClawBench &
Manual pool &
Real web &
${\checkmark}$ &
${\times}$ &
Static release &
153 / 144 sites \\

WildClawBench &
Manual pool &
OpenClaw + sandbox &
${\checkmark}$ &
${\checkmark}$ &
Static release &
60 / 6 cats. \\

PinchBench &
Open-source repos &
OpenClaw stack &
${\checkmark}$ &
Partial &
Static release &
102 \\

Claw-Eval &
Human-verified pool &
Tool/service sandbox &
${\checkmark}$ &
${\checkmark}$ &
Static release &
300 / 9 cats. \\

ResearchClawBench &
Expert-curated pool &
Research workspaces &
${\checkmark}$ &
Partial &
Static release &
40 / 10 domains \\

EvoClaw &
Repo-mined history &
Repo sandbox &
${\checkmark}$ &
${\checkmark}$ &
Static release &
98 / 7 repos \\

LiveCodeBench &
Rolling contests &
Code sandbox &
${\times}$ &
${\checkmark}$ &
Rolling updates &
500+ / 3 plats. \\
\midrule
\textbf{\bench{} (ours)} &
Public signals &
Tool/service sandbox &
${\checkmark}$ &
${\checkmark}$ &
Time-stamped snapshots &
105 / 22 fam. \\
\bottomrule
\end{tabular}
\end{table*}

Relative to these benchmarks, \bench{} is distinguished primarily by how its public release is constructed: the task mix is calibrated to live public workflow signals and materialized as time-stamped benchmark snapshots.
Table~\ref{tab:benchmark_attributes} summarizes this positioning through concrete benchmark attributes and check-style indicators, with task sourcing and release mode distinguishing fixed curated pools from externally refreshed sources such as rolling contests or public workflow signals.
Relative to the neighboring suites, \bench{} differs not merely by drawing from an external source, but by using public workflow signals as the source that feeds benchmark construction.

\section{Benchmark Construction}
\label{sec:construction}

The construction of a workflow-agent benchmark must satisfy two requirements that are often in tension.
It should remain \emph{stable} enough for reproducible comparison, but it should also remain \emph{externally aligned} with the workflows that users currently want agents to automate.
\bench{} addresses this tension by separating benchmark construction into a time-varying signal layer and a fixed release layer.
Public workflow signals determine the distributional prior for a release; executable tasks, fixtures, and graders then materialize that prior into a reproducible benchmark snapshot.
Figure~\ref{fig:liveclaw_overview} summarizes this release path, and Figure~\ref{fig:signal_pipeline} details the signal-to-snapshot transformation.

\subsection{Starting from public workflow signals}

A central design choice in \bench{} is that the task distribution is not specified solely by the benchmark authors.
Instead, we use a public upstream proxy for contemporary workflow demand: the ClawHub Top-500 popular skills, ranked by downloads and popularity at the time of release construction.
These signals are not treated as ground-truth measurements of deployment frequency, economic value, or task difficulty.
Rather, they serve as an externally inspectable prior over which workflow types are currently salient to users.
This distinction matters for benchmark validity: the signals guide the mixture of task families, while the released tasks themselves are still implemented, screened, and graded under controlled evaluation conditions.

This separation gives \bench{} a time-stamped release design without sacrificing reproducibility.
Once a release is published, its tasks, fixtures, and graders remain fixed for stable model comparison.
What changes across releases is the upstream signal snapshot used to construct the next task mixture.
This design lets the benchmark absorb shifts in the workflow ecosystem through future refreshes, rather than by modifying an already published evaluation set.

\subsection{From signal to task: the release-construction pipeline}

\begin{figure*}[t]
\centering
\includegraphics[width=\textwidth]{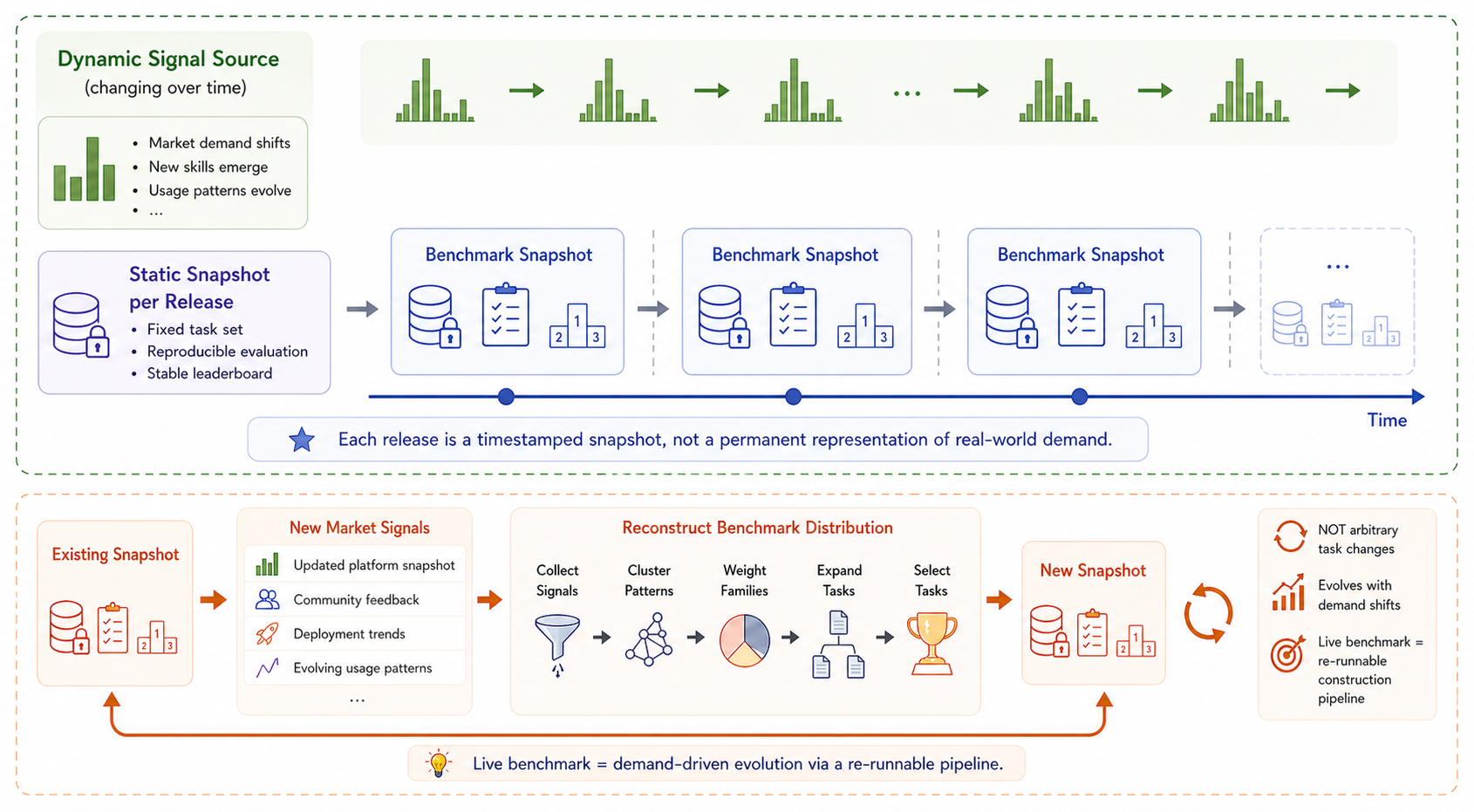}
\caption{Signal-to-snapshot construction in \bench{}.
A refresh starts from public workflow signals, clusters them into stable workflow patterns, converts those patterns into family weights, expands weighted seeds into executable task candidates, and selects a discrimination-aware public subset.
The result is a time-stamped snapshot that remains reproducible for evaluation while staying refreshable when upstream signals drift.}
\label{fig:signal_pipeline}
\end{figure*}

The public signal-to-task pipeline has five conceptual stages, illustrated in Figure~\ref{fig:signal_pipeline}.

\paragraph{Stage 1: Signal collection.}
The pipeline begins from a time-stamped ClawHub Top-500 snapshot rather than from author-written task categories.
Each upstream item is treated as a public signal with provenance, metadata, and a coarse functional label.
This layer provides evidence about which workflows are salient in the surrounding tool ecosystem, but it is not yet an evaluation set: a popular skill name may imply several possible tasks, and a single task may require combining signals from multiple skills.

\paragraph{Stage 2: Pattern clustering.}
Benchmark tasks cannot be defined directly from fragmented skill names, so the next unit is a workflow pattern.
We group related signals by the user goal they imply, the artifact or state they operate on, and the execution surface they require.
The purpose of this step is to remove surface variation in skill names while preserving differences that matter for evaluation, such as whether a workflow is document transformation, cross-tool coordination, data analysis, or workspace repair.
These patterns are then mapped to benchmark-relevant signal families, which serve as mixture units rather than final task labels.

\paragraph{Stage 3: Family weighting.}
Clustered signals are converted into a target release mixture.
Let $r_p$ denote the normalized upstream signal mass of workflow pattern $p$, and let $\mathcal{P}_f$ be the patterns assigned to signal family $f$.
The target family weight is
\[
w_f = \frac{\sum_{p \in \mathcal{P}_f} r_p}{\sum_{p \in \mathcal{P}} r_p}.
\]
These weights are used as a distributional prior for release construction: they influence how many seeds are generated for each family, but they do not by themselves determine whether any individual task is included.

\paragraph{Stage 4: Seed expansion and implementation.}
Weighted patterns are expanded into task seeds that specify the intended user goal, required execution surface, expected evidence, and grading boundary.
Each seed is then materialized as executable candidates with prompts, tool definitions, fixtures, and task-specific graders.
Implementation is followed by pilot screening.
We keep candidates only when they are runnable under the released environment, reproducible under fixed fixtures, and informative enough to produce usable score variation across pilot models.

\paragraph{Stage 5: Discrimination-aware public release selection.}
The public snapshot is not the full screened pool.
It is a selected subset that balances three constraints: release size, family coverage, and leaderboard resolution.
Tasks that are too brittle, too ambiguous, or non-discriminative in pilot runs are excluded, while the selected subset must still cover the intended mixture of workflow families.
The current public release contains \ntasks{} executable tasks, and the selection objective is made explicit by the optimization formulation below.

Family labels are therefore layer-specific.
Signal families define the mixture prior, task families support coverage and selection constraints, and display or analysis groups summarize released results.
We avoid treating any single taxonomy as canonical across all three roles.

From the screened pool, we select the public release using a mixed-integer linear program (MILP) that makes subset construction explicit and reproducible.
Here, ``program'' means a mathematical optimization problem rather than executable code: the task-selection variables are binary, the objective is linear, and the constraints are linear.
Let $\mathcal{T}$ denote the set of \ncandidates{} screened candidates, $\mathcal{M}$ the set of pilot models, and $s_{t,m} \in [0,1]$ the completion score of model $m$ on task $t$.
For each task $t$, let $x_t \in \{0,1\}$ indicate whether $t$ is included in the public release.
Let $N$ be the target release size and $\mathcal{F}$ the set of fine-grained task families.
For a designated pilot ordering of the top-$K$ models $m_1 \succ m_2 \succ \cdots \succ m_K$, define
$p_{t}^{(i,j)} = \mathbf{1}[s_{t,m_i} \ge s_{t,m_j}]$ as the indicator that the ordering $m_i \succ m_j$ is preserved on task $t$.
The MILP objective is:

\begin{equation}
\max_{x} \quad \sum_{t \in \mathcal{T}} \sum_{i < j \le K} p_{t}^{(i,j)} \cdot x_t
\label{eq:milp_obj}
\end{equation}

subject to:
\begin{equation}
\begin{aligned}
\sum_{t \in \mathcal{T}} x_t &= N, &
\sum_{t \in \mathcal{C}_f} x_t &\geq 1 \quad \forall f \in \mathcal{F}, \\
x_t &= 0 \quad \forall t \in \mathcal{Z}, &
x_t &\in \{0,1\} \quad \forall t \in \mathcal{T}.
\end{aligned}
\label{eq:milp_constraints}
\end{equation}

where $\mathcal{C}_f$ is the set of screened candidates in fine-grained family $f$, and $\mathcal{Z}$ is the set of zero-discrimination tasks that all pilot models pass or all pilot models fail under the selection-phase pass rule.
The point of the MILP is not to claim a unique optimal subset, but to turn release size, coverage, and pilot-order preservation into auditable constraints instead of purely manual curation.

\subsection{Execution environments and the released snapshot}
\label{sec:construction_env}

Each released task is a complete executable evaluation unit rather than a prompt alone.
At minimum, a task includes a YAML task definition (\texttt{task.yaml}), task fixtures, tool schemas, and a task-specific grader (\texttt{grader.py}).
The current release spans controlled service workflows and local workspace repair, so tasks can require both cross-service state manipulation and artifact-level changes in a sandboxed workspace.
Table~\ref{tab:release_composition} summarizes the two execution surfaces and the evidence used for grading.
Table~\ref{tab:services} (Appendix~\ref{app:supp}) lists representative controlled services from the public release; each service loads task-specific fixture data, exposes a RESTful API consumed through tool calls, and records an audit log of reads and mutations that later serves as a deterministic evidence source.

\begin{table*}[t]
\centering
\caption{Benchmark composition by execution surface. Each released task is an executable workflow with task-specific fixtures and graders; detailed service and family counts are reported in Appendix~\ref{app:supp}.}
\label{tab:release_composition}
\scriptsize
\renewcommand{\arraystretch}{1.08}
\setlength{\tabcolsep}{3pt}
\begin{tabular}{@{}>{\raggedright\arraybackslash}p{0.16\textwidth} >{\raggedright\arraybackslash}p{0.18\textwidth} >{\raggedright\arraybackslash}p{0.27\textwidth} >{\raggedright\arraybackslash}p{0.25\textwidth} r@{}}
\toprule
\textbf{Execution surface} & \textbf{Representative scope} & \textbf{Required behavior} & \textbf{Grading evidence} & \textbf{Tasks} \\
\midrule
Service-backed workflows &
Business services such as CRM, finance, email, calendar, helpdesk, and knowledge bases &
Retrieve records, coordinate across systems, synthesize decisions, and write controlled state &
Tool traces, service audit logs, ground-truth fixtures, and semantic rubrics when needed &
\nserviceworkflows{} \\
Workspace repair &
Terminal and local workspace tasks, including SHELL and W-family repair cases &
Inspect files and logs, run commands, modify artifacts, and validate the repair &
Command traces, post-run workspace state, generated artifacts, tests, and semantic rubrics when needed &
\nworkspace{} \\
\midrule
\textbf{Total} & & & & \textbf{\ntasks{}} \\
\bottomrule
\end{tabular}
\end{table*}

The end-to-end evaluation flow is as follows.
A task is defined with prompt, tools, and fixtures; executed through controlled services or a workspace loop; logged into machine-readable traces; and finally graded from observable evidence.
This design makes the object of evaluation the whole workflow trajectory rather than the final answer only.

The current snapshot spans \ncats{} fine-grained task families across \ntasks{} released tasks; the four largest raw families (PRODAPP, SHELL, HR, SALES) account for 45 tasks, while the remaining long tail preserves coverage over a wide range of business and repair workflows. The full family-count breakdown is given in Table~\ref{tab:categories} (Appendix~\ref{app:supp}).

\section{Evaluation Setup and Methodology}
\label{sec:method}

\subsection{Current public release and execution protocol}

This paper reports the current public snapshot.
The release contains \ntasks{} tasks and evaluates \nmodels{} public models under a unified protocol.
For workflow slices shown on the public site, the default execution budget is 24 turns and 300 seconds.
Some workspace-repair tasks allow larger budgets when repair and verification require additional command cycles, but all runs follow the same task-specific limits, prompts, tool schemas, and fixed fixtures.
We do not apply model-specific prompt tuning.

Operationally, each evaluation run proceeds as follows:
(i) task fixtures and required services are loaded;
(ii) the agent receives the task prompt and tool definitions;
(iii) the agent interacts with controlled services or a sandboxed workspace;
(iv) the framework records full traces, including tool calls, responses, tokens, wall time, and environment-side artifacts; and
(v) the task-specific grader computes a score in $[0,1]$.
What is preserved is therefore not only the final answer, but the whole execution process that led to it.

\subsection{Evidence sources and recurring grading patterns}
\label{sec:grading}

The public grading summary for \bench{} is \emph{rule-based extraction + structured LLM judging}.
This phrase has two implications.
First, scoring begins from deterministic evidence rather than by delegating the entire decision to an LLM judge.
Second, when LLM judging is used, it is bound to explicit rubrics and grounded in already collected execution evidence.

Across the release, three evidence sources recur:
\begin{itemize}[nosep,leftmargin=1.2em]
\item \textbf{Data Retrieval} (typically 15--20\%): whether the agent called the right tools and inspected the right sources, verified from dispatch logs.
\item \textbf{Data Accuracy} (typically 40--60\%): whether the final entities, numbers, and conclusions match ground-truth fixtures.
\item \textbf{Action Verification} (typically 10--20\%): whether required state changes actually happened, verified from service audit trails or post-run workspace state.
\end{itemize}

Only when these deterministic signals do not fully cover the task does \bench{} add structured LLM judging for semantic dimensions such as completeness, organization quality, or report coherence.

At the implementation level, graders follow three recurring patterns:

\paragraph{Pattern 1: evidence-plus-judge for analytical workflow tasks.}
For analysis-heavy tasks, such as reconciliation, HR review, or business forecasting, graders combine deterministic checks with rubric-bound LLM judging.
Deterministic checks verify tool discipline, entity and numeric correctness, and required writes.
The judge then evaluates semantic dimensions that cannot be reduced to exact-match checks, using task-specific written rubrics.

\paragraph{Pattern 2: operation verification with a smaller semantic component.}
For tasks such as drafting, meeting scheduling, or ticket triage, deterministic verification carries more weight because correctness can often be checked directly from audit logs.
The judge is used only for aspects such as tone, organization, or summary quality when presentation quality remains relevant.

\paragraph{Pattern 3: script-first verification for workspace tasks.}
For SHELL and W-family tasks, where SHELL denotes terminal tasks and W denotes workspace-repair tasks, grading is fully deterministic.
After execution, verification scripts re-check file contents, service health, configuration integrity, or command outputs inside the workspace.
This makes workspace-repair success a property of post-run state rather than self-report or final textual claims.

When a judge is required, we use GPT-5.4 as the judge model.
Each judge call receives the task prompt, the agent trace, structured summaries of observable actions, and the task-specific rubric.
The judge is therefore not asked to hallucinate hidden evidence; it is asked to score semantic dimensions on top of explicit traces and checks.
Because GPT-5.4 is also one of the evaluated models, this design can introduce judge-model bias.
We mitigate this risk by limiting LLM judging to semantic dimensions that deterministic checks cannot cover, grounding every judge input in traces and rubrics, and reporting the deterministic evidence sources used by each task.
We do not treat the judge as an independent human adjudicator.

\subsection{Metrics and ranking rules}

\bench{} reports two primary public metrics.
The first is \textbf{Pass Rate}, the fraction of tasks whose completion score meets the public pass threshold.
The second is \textbf{Overall Completion Score}, the arithmetic mean of task scores across all \ntasks{} tasks, reported on a 0--100 scale.
For this release, the public pass threshold is $\tau=\passthresh{}$.
For model $m$ with task score $s_{t,m} \in [0,1]$ on task $t$, these metrics are:
\[
\mathrm{PassRate}(m)=\frac{1}{|\mathcal{T}|}\sum_{t\in\mathcal{T}}\mathbf{1}[s_{t,m}\ge \tau],
\qquad
\mathrm{Overall}(m)=\frac{100}{|\mathcal{T}|}\sum_{t\in\mathcal{T}}s_{t,m}.
\]

These two metrics play different roles.
Pass Rate emphasizes whether a workflow was completed under the public pass rule.
Overall Completion retains finer-grained information about how close a model came on the tasks it did not fully pass.
Accordingly, models are ranked first by Pass Rate and then by Overall Completion Score.
This ranking rule aims to preserve the distinction between ``can usually finish the workflow'' and ``often gets part of the way there''.

\section{Main Results}
\label{sec:results}

Table~\ref{tab:leaderboard} gives the current public leaderboard used throughout the paper.

\begin{table*}[t]
\centering
\caption{Current public leaderboard for \bench{}.
Models are ranked by Pass Rate; ties are broken by Overall Completion Score.}
\label{tab:leaderboard}
\small
\begin{tabular}{clllcr}
\toprule
\textbf{\#} & \textbf{Model} & \textbf{Organisation} & \textbf{Pass Rate} & \textbf{Pass Count} & \textbf{Overall} \\
\midrule
 1 & Claude Opus 4.6    & Anthropic   & 66.7\% & 70 & 83.6 \\
\rowcolor{tablegray}
 2 & GPT-5.4            & OpenAI      & 63.8\% & 67 & 81.7 \\
 3 & Claude Sonnet 4.6  & Anthropic   & 61.9\% & 65 & 79.9 \\
\rowcolor{tablegray}
 4 & GLM-5              & Zhipu AI    & 61.9\% & 65 & 78.1 \\
 5 & MiniMax M2.7       & MiniMax     & 54.3\% & 57 & 77.5 \\
\rowcolor{tablegray}
 6 & MiMo V2 Pro        & Xiaomi      & 53.3\% & 56 & 76.9 \\
 7 & Kimi K2.5          & Moonshot AI & 53.3\% & 56 & 76.2 \\
\rowcolor{tablegray}
 8 & Gemini 3.1 Pro     & Google      & 53.3\% & 56 & 74.0 \\
 9 & DeepSeek V3.2      & DeepSeek    & 51.4\% & 54 & 69.3 \\
\rowcolor{tablegray}
10 & Qwen 3.6 Plus      & Alibaba     & 50.5\% & 53 & 71.4 \\
11 & MiniMax M2.5       & MiniMax     & 50.5\% & 53 & 70.9 \\
\rowcolor{tablegray}
12 & Qwen 3.5 397B      & Alibaba     & 49.5\% & 52 & 72.7 \\
13 & Doubao Seed 2.0    & ByteDance   & 43.8\% & 46 & 70.4 \\
\bottomrule
\end{tabular}
\end{table*}

\subsection{The most direct takeaway from the current leaderboard}
\label{sec:results_main}

The most direct takeaway from the current public leaderboard is that the ceiling is still far away.
The top model, Claude Opus~4.6, reaches 66.7\% pass rate with an overall completion score of 83.6, and GPT-5.4 follows at 63.8\% and 81.7.
In other words, even the best publicly reported model does not cross the 70\% pass-rate mark.
This indicates that stable workflow execution remains meaningfully short of reliable automation.

\begin{figure}[t]
\centering
\includegraphics[width=0.94\textwidth]{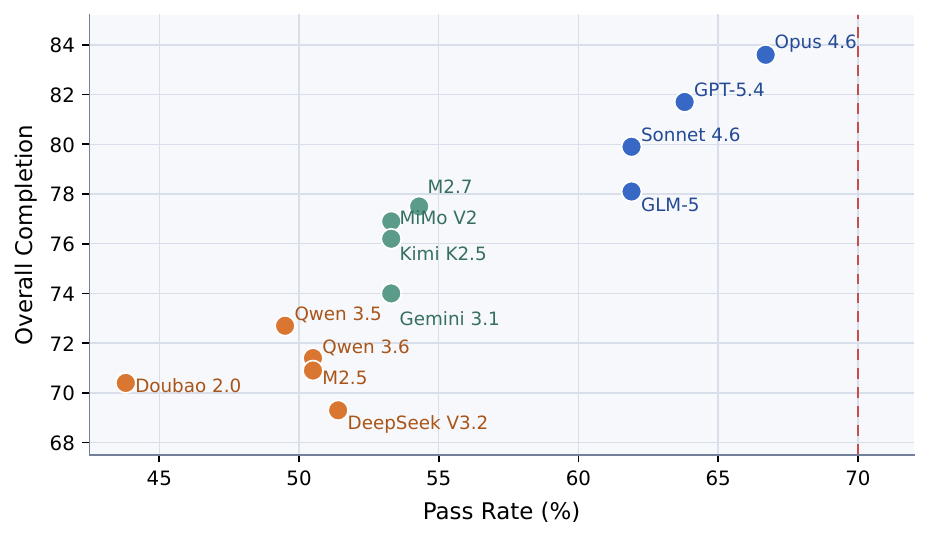}
\caption{Leaderboard metric landscape for \bench{}.
Each point is one model, with Pass Rate on the x-axis and Overall Completion on the y-axis.
The dashed line marks the 70\% pass-rate level: no model crosses it, while models with similar pass rates can still differ in overall completion.
The complete numeric ranking is given in Table~\ref{tab:leaderboard}.}
\label{fig:model_landscape}
\end{figure}

The leaderboard is also not tightly clustered.
The gap between the top and bottom models is 22.9 percentage points in pass rate.
At the same time, several clusters show that pass rate alone is not enough:
MiMo V2 Pro, Kimi K2.5, and Gemini 3.1 Pro all tie at 53.3\% pass rate but are separated by overall completion, indicating that the benchmark captures meaningful differences in how completely models solve partially successful tasks.

\subsection{Family-level heterogeneity}
\label{sec:results_families}

\begin{figure*}[t]
\centering
\includegraphics[width=\textwidth]{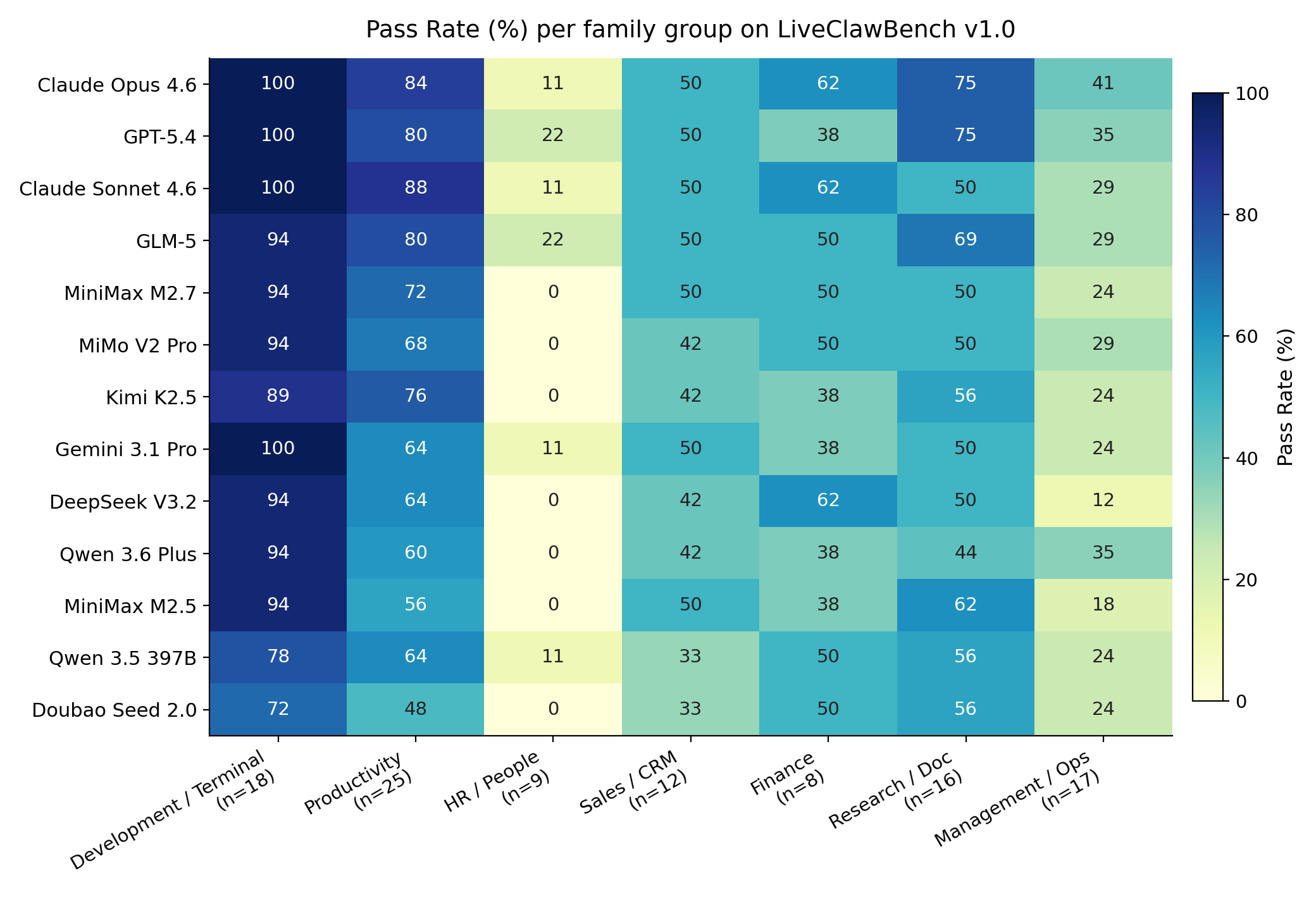}
\caption{Pass-rate heatmap for \nmodels{} models over seven analysis groups derived from the released task families.
Group sizes are: Development / Terminal (\nworkspace{} tasks), Productivity (25), HR / People (9), Sales / CRM (12), Finance (8), Research / Doc (16), and Management / Ops (17).
Development / Terminal is near-ceiling for the strongest models, Productivity shows the largest spread, and HR / People remains difficult for all models.}
\label{fig:heatmap}
\end{figure*}

Aggregate rankings hide a large amount of family-level structure.
Figure~\ref{fig:heatmap} groups the release into seven analysis buckets for readability.
Two patterns are immediately visible.
First, Development / Terminal tasks are comparatively easy for current frontier models.
Second, business-facing analytical and coordination workflows remain much harder and much more heterogeneous.

The grouped heatmap unpacks into consistent fine-grained behaviour.
Development / Terminal is near ceiling: Claude Opus~4.6, GPT-5.4, and Claude Sonnet~4.6 all reach 100\% on the grouped slice, and even the lowest model remains above 72\%.
By contrast, HR / People is extremely hard: no model is above 22.2\%, and several models score 0.0\%.
The Productivity slice has the widest spread, from 88.0\% for Claude Sonnet~4.6 to 48.0\% for Doubao Seed~2.0, showing that task types involving planning, coordination, and multi-step productivity workflows remain strongly differentiating.

At the fine-grained family level, PRODAPP averages 84.2\% pass rate across models but still exhibits a 47.1-point spread between the best and worst models.
HR averages only 6.8\%.
MGMT is all-fail under the public pass rule, and WORKFLOW averages 12.8\%.
Some of the strongest discriminators in the release are the meeting-prep action task (\texttt{CTB\_COMM\_24}), multi-document merge (\texttt{CTB\_D01}), e-commerce monthly reconcile (\texttt{CTB\_DATA\_08}), and first-response-time audit (\texttt{CTB\_SUPPORT\_03}).
These are exactly the kinds of workflows where missing one source, one entity link, or one action can collapse the final score.

\subsection{Service-backed workflows versus workspace repair}

The release splits into two execution surfaces: service-backed workflows (\nserviceworkflows{} tasks) and terminal/workspace repair (\nworkspace{} tasks, i.e.\ SHELL + W).
The asymmetry is stark.
All models score at least 72.2\% pass rate on the workspace slice, and several models are at or near 100\%.
But no model is above 59.8\% on service-backed workflows.
Claude Opus~4.6 leads that slice at 59.8\%, with GPT-5.4 at 56.3\% and GLM-5 at 55.2\%.

This matters because it changes how the leaderboard should be interpreted.
Current agents are often already competent at local diagnosis and repair under fixed constraints.
The harder bottleneck is coordinated, evidence-sensitive business execution across multiple services.
In other words, the main gap is not ``can the model use a terminal at all?'' but ``can it complete a multi-system workflow without losing state, missing evidence, or failing to execute required writes?''

\subsection{Ranking divergence and what \bench{} is really measuring}

The ranking induced by \bench{} does not simply mirror general chat or writing ability.
That is expected.
The benchmark does not reward only the readability of the final answer; it also rewards cross-system evidence collection, correct record linkage, action completion, and post-run state integrity.
A model that writes fluent summaries can therefore still score poorly if it failed to call the required tools, missed crucial evidence, or left the workspace in the wrong state.

This is the core ``can say'' versus ``can do'' distinction.
For example, onboarding-style HR tasks often elicit polished but generic writeups.
However, under action-grounded grading, these outputs fail if they omit employee-specific details, miss required tool calls, or do not satisfy the task's evidence checks.
The same pattern appears in management and workflow families, where the final text may look plausible while the underlying evidence trail remains incomplete.
\bench{} therefore measures not just linguistic plausibility, but executional closure under traceable evidence.

\subsection{Threshold effects and task discrimination}

\begin{figure*}[t]
\centering
\makebox[\textwidth][c]{\includegraphics[width=1.08\textwidth]{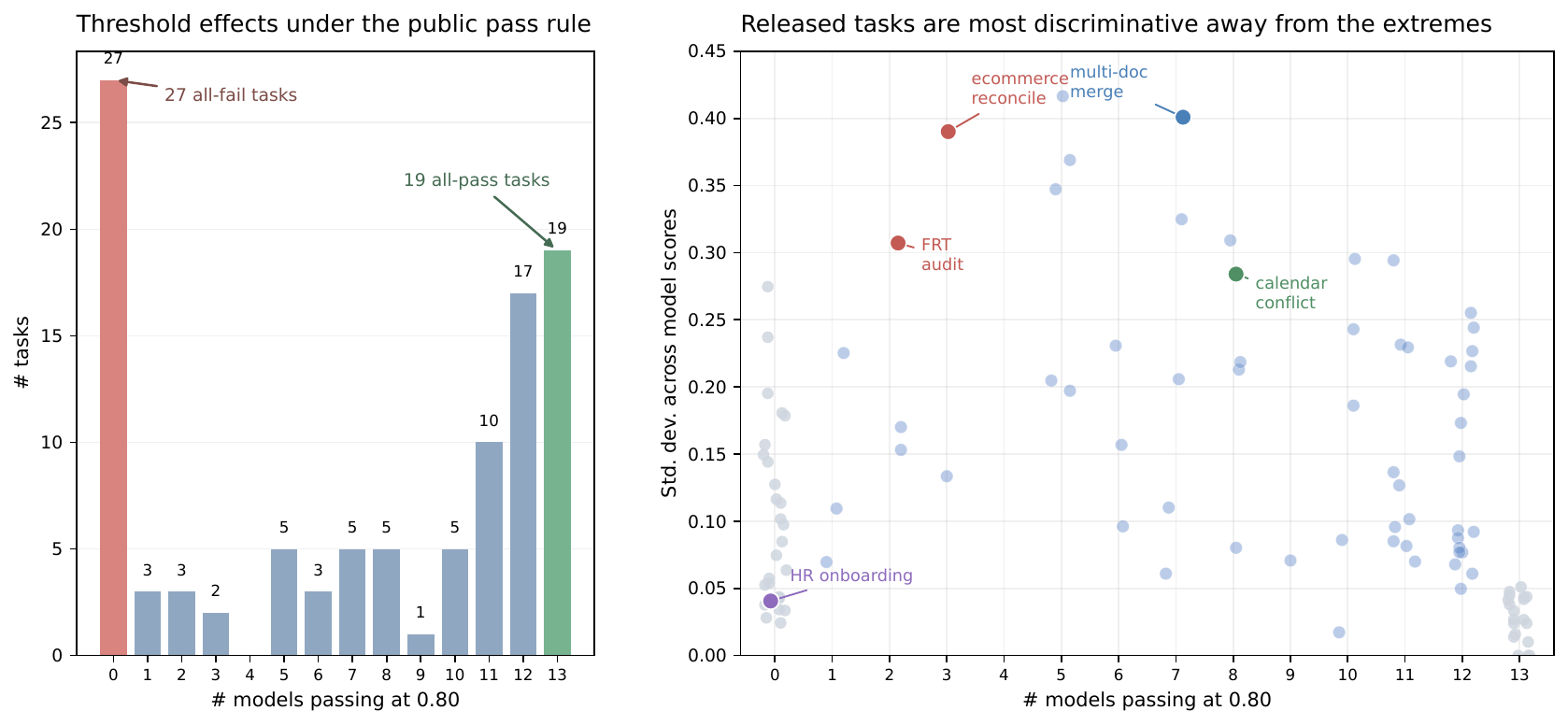}}
\caption{Threshold-aware task discrimination in the public release. Under the public pass rule, tasks split into all-fail and all-pass clusters, while discrimination concentrates in the middle band. This is why selection uses a more permissive pilot-time rule even though evaluation reports the stricter public criterion.}
\label{fig:discrimination}
\end{figure*}

Not all released tasks contribute equally to model separation.
We measure task discrimination as the standard deviation of completion scores across the \nmodels{} leaderboard models.
Figure~\ref{fig:discrimination} makes the threshold structure explicit: many released tasks collapse to the extremes under the public pass rule, while the most discriminative tasks sit in the middle of the pass-count range.
The strongest discriminators include \texttt{ecommerce\_monthly\_reconcile}, \texttt{first\_response\_time\_audit}, and \texttt{multi\_doc\_merge}, all of which require precise multi-source extraction where partial tool usage leads to large score drops.
Conversely, the least discriminating released tasks are predominantly SHELL or W tasks where most models already perform near ceiling.

An important subtlety is that public discrimination is measured under the public pass rule, whereas selection-time filtering uses a more permissive pilot-time rule.
Under this public rule, 19 released tasks are all-pass and 27 are all-fail across the 13 public leaderboard models.
This does not invalidate the curation procedure; it reflects the fact that the public leaderboard uses a stricter success criterion while still enforcing broad family coverage.
For this reason, pass rate should be interpreted together with overall completion rather than in isolation.

\subsection{Efficiency and resource use}

Accuracy is only part of a deployment decision.
Resource use matters as well.
Table~\ref{tab:efficiency} reports release-level resource totals for all \nmodels{} public leaderboard models.
Cost values are estimated API costs from recorded input/output token use and release-time provider list prices, not billed spend or full experiment cost.

\begin{table}[t]
\centering
\caption{Efficiency metrics for all public leaderboard models, aggregated over the \ntasks{} public tasks.
Cost totals are estimated API costs from recorded input/output token usage and release-time provider list prices.}
\label{tab:efficiency}
\small
\setlength{\tabcolsep}{5pt}
\renewcommand{\arraystretch}{0.95}
\begin{tabular}{lrrrr}
\toprule
\textbf{Model} & \textbf{Tokens (M)} & \textbf{Turns} & \textbf{Est. API Cost} & \textbf{Time} \\
\midrule
Claude Opus~4.6    & 3.32 & 506 & \$31.61 & 213 min \\
\rowcolor{tablegray}
GPT-5.4            & 1.26 & 373 & \$6.27  & 104 min \\
Claude Sonnet~4.6  & 2.41 & 417 & \$14.35 & 241 min \\
\rowcolor{tablegray}
GLM-5              & 1.70 & 424 & \$2.46  & 169 min \\
MiniMax M2.7       & 1.50 & 395 & \$0.69  & 102 min \\
\rowcolor{tablegray}
MiMo V2 Pro        & 1.58 & 371 & \$2.17  & 103 min \\
Kimi K2.5          & 1.61 & 449 & \$1.06  & 205 min \\
\rowcolor{tablegray}
Gemini 3.1 Pro     & 3.39 & 518 & \$12.08 & 172 min \\
DeepSeek V3.2      & 2.06 & 617 & \$0.56  & 118 min \\
\rowcolor{tablegray}
Qwen 3.6 Plus      & 1.42 & 366 & \$1.41  & 216 min \\
MiniMax M2.5       & 1.81 & 465 & \$0.78  & 79 min  \\
\rowcolor{tablegray}
Qwen 3.5 397B      & 2.73 & 490 & \$1.92  & 146 min \\
Doubao Seed 2.0    & 1.86 & 474 & \$1.47  & 141 min \\
\bottomrule
\end{tabular}
\end{table}

Among the top-ranked models, GPT-5.4 offers the best efficiency profile: it uses the fewest tokens among the top four models, is faster than the other top four models, and remains close to the top of the leaderboard.
Claude Opus~4.6 is the most accurate model but has materially higher estimated API cost.
Lower-cost models such as MiniMax M2.5, MiniMax M2.7, DeepSeek V3.2, and Kimi K2.5 still trail the top models on pass rate.
This suggests that deployment choices for workflow agents should be made against family-level accuracy and cost constraints rather than against a single aggregate ranking alone in practice.

\section{Conclusion}
\label{sec:conclusion}

We introduced \bench{}, a live agent benchmark built around two design goals: workflow calibration and evidence alignment.
Workflow calibration keeps the released task mix close to workflows users currently want to automate.
Evidence alignment grounds scores in what the agent observably did rather than in final-text plausibility alone.
The current snapshot shows both why this matters and where current agents fall short.
The best model passes only 66.7\% of tasks, local workspace repair is materially easier than API-backed business workflows, and HR, management, and multi-system coordination remain far from solved.
Each release is therefore best read as a time-stamped public snapshot rather than a permanent definition of workflow-agent ability.
Across releases, the signal layer can be refreshed from new public workflow signals, allowing the benchmark to track changing automation demand without making past evaluations unstable.
The long-term value of the benchmark is to make every release explicit about why it was constructed, what it measures well, and what it still leaves out.

\bibliography{references}
\bibliographystyle{abbrvnat}

\clearpage
\appendix

\section{Supplementary Figures and Tables}
\label{app:supp}

\begin{figure}[ht]
\centering
\includegraphics[width=\textwidth]{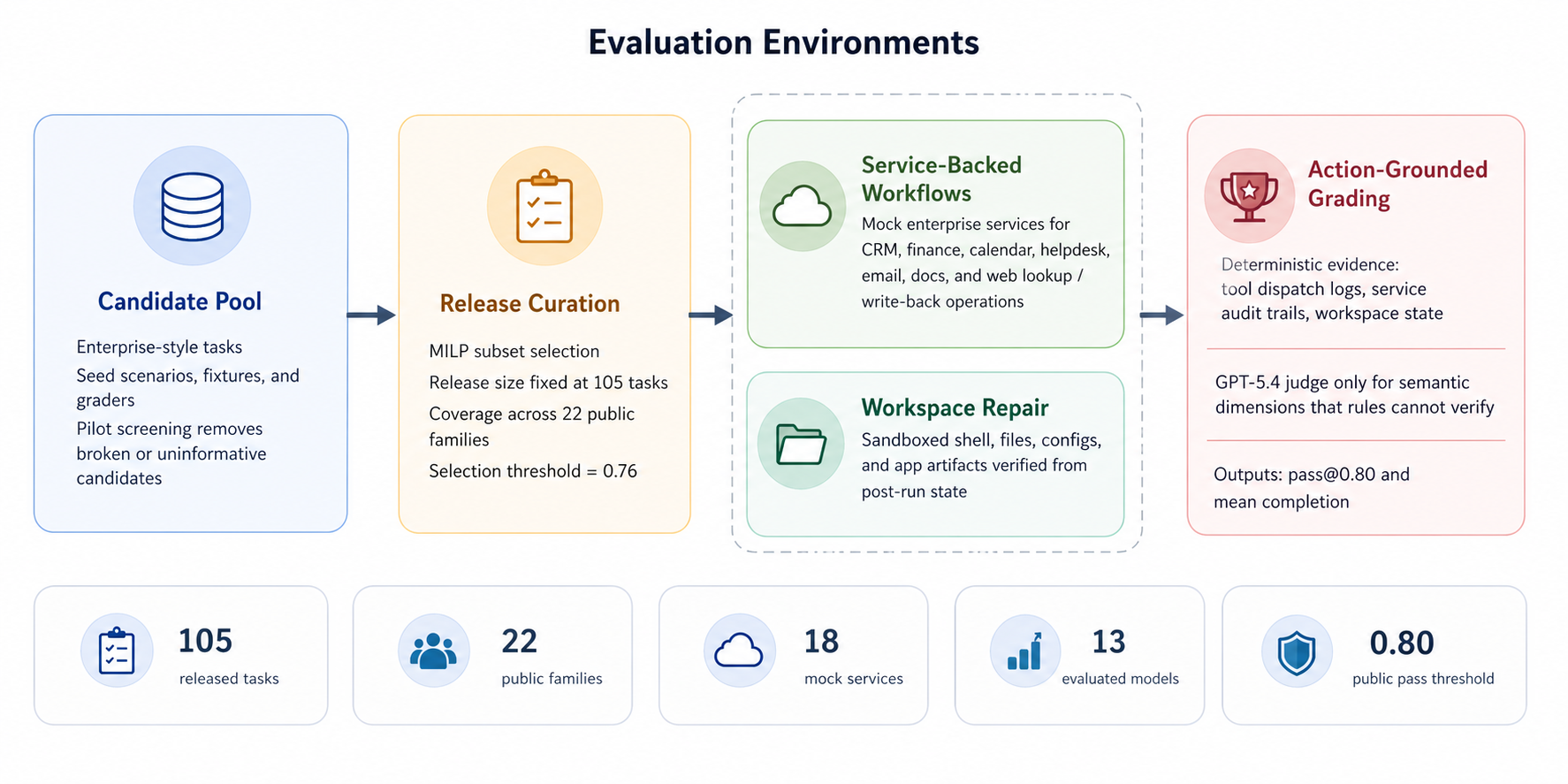}
\caption{Overview of \bench{}. Candidate tasks are piloted and screened, curated into a \ntasks{}-task public release with family-coverage constraints, executed in either service-backed workflow or workspace-repair environments, and graded from observable execution evidence. GPT-5.4 is used only for semantic dimensions that deterministic checks cannot verify.}
\label{fig:overview}
\end{figure}

This appendix collects construction- and setup-reference material that accompanies the main text.
Figure~\ref{fig:overview} gives the benchmark-construction overview; Table~\ref{tab:services} lists representative controlled services; Table~\ref{tab:categories} reports the released family counts.

\begin{table}[ht]
\centering
\caption{Representative controlled services in \bench{}. The public release additionally includes controlled variants for web retrieval, media handling, and robustness testing.}
\label{tab:services}
\small
\begin{tabular}{ll}
\toprule
\textbf{Service} & \textbf{Capability} \\
\midrule
\texttt{calendar}  & Event scheduling and conflict detection \\
\texttt{contacts}  & Directory and address-book lookup \\
\texttt{crm}       & Customer and employee record management \\
\texttt{documents} & File and document storage \\
\texttt{finance}   & Transactions, budgets, reconciliation data \\
\texttt{gmail}     & Email listing, reading, drafting, sending \\
\texttt{helpdesk}  & Support ticket management and triage \\
\texttt{kb}        & Knowledge base articles \\
\texttt{notes}     & Document and note CRUD operations \\
\texttt{scheduler} & Job scheduling and cron management \\
\texttt{todo}      & Task list management \\
\texttt{web}       & Controlled web content retrieval \\
\bottomrule
\end{tabular}
\end{table}

\begin{table}[ht]
\centering
\caption{Public leaderboard task-family distribution in \bench{}.}
\label{tab:categories}
\small
\begin{tabular}{lrlr}
\toprule
\textbf{Family} & \textbf{Tasks} & \textbf{Family} & \textbf{Tasks} \\
\midrule
A         & 4  & MGMT      & 4  \\
C         & 2  & OPS       & 3  \\
COMM      & 5  & PROD      & 4  \\
CRM       & 4  & PRODAPP   & 17 \\
D         & 3  & R         & 1  \\
DATA      & 5  & RESEARCH  & 2  \\
DOC       & 1  & SALES     & 6  \\
FIN       & 4  & SEC       & 1  \\
HR        & 9  & SHELL     & 13 \\
IR        & 4  & SUPPORT   & 5  \\
W         & 5  & WORKFLOW  & 3  \\
\midrule
\multicolumn{3}{l}{\textbf{Total}} & \textbf{\ntasks{}} \\
\bottomrule
\end{tabular}
\end{table}

\section{Full Score Matrix}
\label{app:matrix}

Table~\ref{tab:full_matrix} presents a representative subset of the task $\times$ model score matrix.
The full \ntasks{} $\times$ \nmodels{} matrix is available in the project release as \texttt{benchmark/results/raw\_matrix\_v3.csv}.

\begin{table}[ht]
\centering
\caption{Representative subset of the score matrix (20 tasks $\times$ 6 models).
Scores that meet the public pass rule are shown in \textbf{bold}.}
\label{tab:full_matrix}
\small
\setlength{\tabcolsep}{3pt}
\begin{tabular}{l*{6}{r}}
\toprule
\textbf{Task ID} & \textbf{Opus} & \textbf{GPT-5.4} & \textbf{Sonnet} & \textbf{GLM-5} & \textbf{Kimi} & \textbf{DS V3.2} \\
\midrule
A01\_financial\_recon.        & \textbf{0.94} & 0.62 & \textbf{0.94} & \textbf{0.91} & 0.74 & \textbf{0.90} \\
COMM\_24\_meeting\_prep       & 0.05 & 0.76 & \textbf{0.91} & 0.05 & \textbf{0.91} & 0.71 \\
DATA\_08\_ecommerce\_recon.   & \textbf{0.94} & \textbf{0.90} & 0.07 & \textbf{0.80} & 0.00 & 0.02 \\
D01\_multi\_doc\_merge        & \textbf{0.90} & \textbf{0.94} & 0.00 & \textbf{0.92} & 0.56 & 0.78 \\
HR\_01\_onboarding            & 0.56 & 0.53 & 0.55 & 0.42 & 0.53 & 0.53 \\
PRODAPP\_01\_cal\_conflict    & \textbf{0.99} & \textbf{0.94} & \textbf{0.99} & \textbf{0.94} & \textbf{0.95} & 0.42 \\
PRODAPP\_15\_capacity\_fc.    & 0.65 & \textbf{0.94} & \textbf{0.91} & 0.26 & \textbf{0.88} & \textbf{0.85} \\
R03\_whiteboard\_platform     & \textbf{0.98} & \textbf{0.98} & 0.05 & \textbf{0.82} & \textbf{0.98} & \textbf{0.90} \\
SHELL\_03\_disk\_usage        & \textbf{1.00} & \textbf{1.00} & \textbf{1.00} & \textbf{1.00} & \textbf{1.00} & \textbf{1.00} \\
SHELL\_22\_cache\_hit         & \textbf{0.91} & \textbf{0.96} & \textbf{0.87} & \textbf{0.90} & \textbf{0.87} & \textbf{0.96} \\
SUPPORT\_03\_frt\_audit       & \textbf{0.81} & 0.70 & 0.60 & \textbf{0.82} & 0.47 & 0.10 \\
W04\_devops\_deploy           & \textbf{1.00} & \textbf{1.00} & \textbf{1.00} & \textbf{1.00} & \textbf{1.00} & \textbf{1.00} \\
WORKFLOW\_02\_expense\_appr.  & \textbf{0.80} & 0.41 & 0.48 & 0.54 & 0.58 & \textbf{0.80} \\
FIN\_29\_budget\_reforecast   & \textbf{0.82} & 0.19 & \textbf{0.82} & 0.19 & 0.57 & \textbf{0.83} \\
SALES\_10\_key\_account       & \textbf{0.99} & \textbf{0.97} & \textbf{0.92} & \textbf{0.94} & 0.05 & \textbf{0.98} \\
SEC\_01\_suspicious\_login    & \textbf{0.93} & \textbf{0.98} & \textbf{0.99} & \textbf{0.98} & \textbf{0.93} & \textbf{0.94} \\
IR\_14\_esg\_compliance       & \textbf{0.98} & \textbf{0.89} & \textbf{0.98} & \textbf{0.85} & \textbf{0.99} & \textbf{0.98} \\
CRM\_04\_lead\_scoring        & \textbf{0.89} & \textbf{0.90} & \textbf{0.98} & \textbf{0.89} & \textbf{0.89} & \textbf{0.98} \\
OPS\_04\_dependency\_map      & 0.75 & \textbf{0.85} & 0.79 & 0.64 & 0.78 & 0.69 \\
MGMT\_01\_okr\_review         & 0.51 & 0.49 & 0.44 & 0.47 & 0.07 & 0.49 \\
\bottomrule
\end{tabular}
\end{table}

\section{Grader Architecture Details}
\label{app:grader}

All task graders inherit from \texttt{AbstractGrader}, which provides:
\begin{itemize}[nosep,leftmargin=1.2em]
\item \texttt{\_tool\_gate(dispatches)}: computes a multiplicative penalty based on whether the agent called the required APIs.
\item \texttt{compute\_robustness(dispatches)}: computes the fraction of tool calls that returned HTTP 2xx status.
\item \texttt{format\_conversation(messages)}: formats the full agent trace for judge input.
\item \texttt{summarize\_actions(audit\_data)}: extracts observable actions from service audit logs.
\item \texttt{\_get\_all\_assistant\_text(messages)}: concatenates assistant responses for keyword matching when needed.
\end{itemize}

The grader registry (\texttt{registry.py}) dynamically loads each task's \texttt{grader.py} at scoring time using Python \texttt{importlib}, locating the \texttt{AbstractGrader} subclass by reflection.
This preserves a shared grading interface while allowing task-specific evidence logic where necessary.

\section{Mixed-Integer Linear Programming Solver Details}
\label{app:milp}

Mixed-integer linear programming (MILP) is the optimization form used for public-subset selection: some decision variables are integer-valued, here binary task-inclusion indicators, while the objective and constraints remain linear.
The MILP defined in Eqs.~\ref{eq:milp_obj}--\ref{eq:milp_constraints} is solved with a standard branch-and-bound solver.
The problem contains \ncandidates{} binary variables, \ncats{} family-coverage constraints, and $O(K^2 \cdot |\mathcal{T}|)$ pairwise-order terms in the objective, where $K=5$ for the pilot ordering panel.
At this scale, the optimization is readily tractable with off-the-shelf MILP solvers.

The pilot ordering used during selection is Claude Opus $\succ$ GPT-5.4 $\succ$ Claude Sonnet $\succ$ GLM-5 $\succ$ Kimi K2.5, derived from preliminary evaluation on the screened candidate pool.
This target ordering is a practical release-construction device rather than a claim about a permanent global ranking.

\end{document}